\documentclass[useAMS,usenatbib]{mn2e}
\usepackage{times}
\usepackage{aas_macros}
\usepackage{amssymb}
\usepackage{graphicx}
\usepackage{rotating}
\usepackage{lscape}
\usepackage{multirow}

\def\sfrd{$\rho_{\rm SFR}$\ }

\begin{document}

\title[H$\alpha$ emitters at $z=2.23$]{HiZELS: a high redshift survey
  of H$\alpha$ emitters. I: the cosmic star-formation rate 
  and clustering at $\mathbf{z=2.23}$\thanks{Based on observations
obtained with the Wide Field CAMera (WFCAM) on the United Kingdom
Infrared Telescope (UKIRT)}} 
\author[J. E. Geach et al.]
{\parbox[h]{\textwidth}{
J.\ E.\ Geach$^1$\thanks{E-mail:
j.e.geach@durham.ac.uk}, Ian\ Smail$^1$, P.\ N.\ Best$^2$, J.\ Kurk$^3$,
M.\ Casali$^4$, R.\ J.\ Ivison$^{2,5}$ \&
K.\ Coppin$^1$}
\vspace*{6pt}\\
\noindent $^1$Institute for Computational Cosmology, Durham University, South Road, 
Durham. DH1 3LE. UK.\\
\noindent $^2$SUPA, Institute for Astronomy, Royal Observatory, University
of Edinburgh, Blackford Hill, Edinburgh. EH9 3HJ. UK. \\
\noindent $^3$Max-Planck-Institut f\"{u}r Astronomie, K\"{o}nigstuhl,
17 D--69117, Heidelberg, Germany\\
\noindent $^4$European Southern Observatory,
Karl-Schwarzschild-Strasse 2, D--85738 Garching, Germany\\
\noindent $^5$Astronomy Technology Centre, Royal Observatory,
University of Edinburgh, Blackford Hill, Edinburgh. EH9 3HJ. UK. }

\date{}

\pagerange{\pageref{firstpage}--\pageref{lastpage}}\pubyear{2008}

\maketitle

\label{firstpage}

\begin{abstract}
We present results from a near-infrared narrow-band survey of emission-line
galaxies at $z=2.23$, using the Wide Field Camera on the
United Kingdom Infrared Telescope. The H$_2$S1 narrow-band filter
($\lambda_c=2.121\mu$m) we employ selects  the 
H$\alpha$ emission line redshifted to $z=2.23$, and is
thus suitable for selecting 
`typical' star forming galaxies and active galactic nuclei at this
epoch. The pilot study was undertaken in the well
studied Cosmological Evolution Survey field (COSMOS) and is already
the largest near-infrared narrow-band survey at this depth, with a
line flux limit of $F_{\rm H\alpha}\sim10^{-16}$\,erg\,s$^{-1}$\,cm$^{-2}$
over 0.60\,square\,degrees, probing
$\sim$$220\times 10^3$ Mpc$^3$ (co-moving) down to a limiting star
formation rate of $\sim$30\,$M_\odot$\,yr$^{-1}$ (3$\sigma$). In this paper we
present the results from our pilot survey and evaluate the H$\alpha$ luminosity
function and estimate the clustering properties of H$\alpha$ emitters
at $z=2.23$ from 55 detected galaxies. The integrated
luminosity function is used to estimate the volume averaged star
formation rate at $z=2.23$: $\rho_{\rm SFR} = 0.17^{+0.16}_{-0.09}$\,$M_\odot\,{\rm yr^{-1}}\,{\rm
  Mpc^{-3}}$ for $L_{\rm H\alpha} > 10^{42}$\,erg\,s$^{-1}$. For the
first time, we use the H$\alpha$ star-formation tracer to reliably constrain
$\rho_{\rm SFR}$ out to $z=2.23$ demonstrating the rapid increase in
$\rho_{\rm SFR}$ out to this redshift as well as confirming
the flattening in $\rho_{\rm SFR}$ between $z\sim1$--2.
In addition to the luminosity distribution, we analyse the clustering
properties of these galaxies. 
Using the 2-point angular correlation function, $\omega(\theta)$, we 
estimate a real space correlation
length of $r_0= 4.2^{+0.4}_{-0.2}$\,$h^{-1}$\,Mpc. In comparison to models of
clustering which take into account bias evolution, we estimate that
these galaxies are hosted by dark matter halos of mass $M_{\rm halo} \sim
10^{12}\,M_\odot$ consistent with the progenitors of the Milky Way.
\end{abstract}

\def\sfrd{$\rho_{\rm SFR}$\ }

\section{Introduction}
One of the most fundamental, and challenging, goals of modern observational
cosmology is to reliably determine the volume-averaged star-formation history 
of the Universe: its distribution function and variation with environment are 
powerful tools for understanding the physics of galaxy formation and evolution. 
Surveys using a range of star-formation indicators suggest that the star-formation rate
density ($\rho_{\rm SFR}$) rises as $\sim$$(1+z)^{4}$ out to $z\sim1$ (e.g.\
Lilly et al.\ 1995) and hence the `epoch' of galaxy formation must
occur at $z>1$.  Unfortunately, 
tying down the precise epoch of maximum activity at $z>1$ is more difficult, with 
different star-formation indicators giving a wide spread in $\rho_{\rm SFR}$
(e.g. Smail et al.\ 2002; Hopkins 2004) -- although there is a general
trend for $\rho_{\rm SFR}$ to plateau at $z\sim2$ and hence it is likely
that many of the stars seen in galaxies at the present-day were formed
between $z\sim1$--$3$.  Indeed, $z\sim 2$--$2.5$ appears to be a critical
era in the evolution of many populations: optical QSO activity (which
should be linked to the growth of super-massive black holes) peaks at
this epoch (Boyle et al.\ 2000), as does the luminosity density in
bright submillimetre galaxies -- thought to represent an early phase in the
formation of massive galaxies (Chapman et al.\ 2005).  This behaviour
might be associated with the formation of 
spheroids or the most massive galaxies. It is not clear
that this evolution is true of 
``typical'' star forming galaxies (i.e. those with SFRs
$<$100\,$M_\odot$\,yr$^{-1}$). A sensitive survey of star forming
galaxies in a range of representative environments at high-$z$ is needed to address
this issue.

Variation between the different $\rho_{\rm SFR}$ measurements at high-$z$ is in part due to a
combination of the effects of sample selection, Cosmic variance and biases in
individual indicators. Estimates from ultra-violet--selected samples
(e.g.\ Madau et al.\ 1996) require large corrections for dust
extinction and may miss the most obscured sources
altogether. Similarly, although far-infrared-selected samples are less
influenced by  
dust extinction, they require large extrapolations to account for
sources below their bright luminosity limits (e.g.\ Chapman et al.\ 2005).
Thus while no single indicator can give an unbiased view of the
evolution of the $\rho_{\rm SFR}$, mixing different indicators at different
epochs is not helping our understanding.

What we require is one star-formation indicator that can be
applied from $z=0$--3, is relatively immune to dust extinction
and which has sufficient sensitivity that $\rho_{\rm SFR}$ estimates do not
require large extrapolations for faint sources. The H$\alpha$ emission
line luminosity is a well-calibrated SFR indicator providing  a good measure of
the instantaneous SFR of  young, massive
stars ($>$8\,$M_\odot$; Kennicutt 1998). Thus H$\alpha$ combines all of these properties and
hence is the tool of choice for studying the evolution of the SFR
density out to $z\sim3$ -- covering the expected peak in the
SFRD. 
The H$\alpha$ luminosity function (LF), and the corresponding $\rho_{\rm SFR}$ has
been measured out to to $z\sim1.3$ by a series of investigators
(e.g. Gallego et al.\ 1995; Tresse et al.\ 2002; Yan et al.\ 1999;
Shioya et al. 2008; Villar et al. 2008).  These
observations confirm the rapid rise seen by other indicators out to
$z\sim1$ (see also the compilation of Moorwood 2004). 

At higher redshift attempts to search for H$\alpha$ emission
have been made with near-infrared narrow-band surveys (e.g. Thompson et al.\
1996; Moorwood et al.\ 2000), but these have been hampered by
small survey areas. 
At the typical limit of these studies, $F_{\rm
  H\alpha}\sim 10^{-16}$\,erg\,s$^{-1}$\,cm$^{-2}$, only a handful of potential
$z\sim2$ H$\alpha$ emitters have been detected. Indeed, in the largest
survey of this type to date, Thompson et al. (1996) detected only {\it
  one} H$\alpha$ emitter over 276\,square arcminutes to a flux limit of
$F_{\rm H\alpha}  = 3.5\times
10^{-16}$\,erg\,s$^{-1}$\,cm$^{-2}$. This was confirmed to have an
emission line corresponding to $z=2.43$ in the follow-up spectroscopic
survey of Beckwith et al.\ (1998). In comparison, Moorwood et
al. (2000) probe to a slightly deeper limit, $F_{\rm H\alpha}  = 0.5\times
10^{-16}$\,erg\,s$^{-1}$\,cm$^{-2}$ over 100\,square arcminutes, and detect
10 candidate emitters, but all of the candidates chosen for
spectroscopic follow-up turned out to be higher redshift [O{\sc
  iii}]$\lambda$5007 emitters (Kurk et al. 2004; Moorwood et
al. 2003). Detecting large
numbers of H$\alpha$ emitters using this method is clearly
challenging, and larger (and preferably deeper) surveys are needed to
efficiently probe the H$\alpha$ luminosity density at high redshift.

Larger scale narrow-band imaging surveys would also present the
opportunity to measure the
clustering properties of high-$z$ star forming galaxy populations. Not
only does this provide
an important insight into the real space distribution of galaxies, but
(assuming a model for bias -- i.e. how the observed galaxy populations
trace the underlying matter distribution) can provide an estimate of the
mass of dark matter halos that host such galaxies. Comparing the
measurements of clustering for populations at high and low redshift, it
is therefore possible to infer the likely progenitor populations of galaxies seen
in the local Universe. 

In this paper we make an improvement over previous $z\sim2$ H$\alpha$
surveys by utilising the panoramic imaging capability of the Wide
Field Camera (WFCAM, Casali et al. 2007) on the 3.8-m United Kingdom Infrared Telescope
(UKIRT). We have obtained deep narrow-band imaging at $\lambda=2.121$$\mu$m
to search for $z=2.23$ H$\alpha$ emitters over 0.603\,sq. degrees
of the Cosmological Evolution Survey field (COSMOS) (Scoville et al.\
2007). This corresponds to a co-moving volume $\sim$30$\times$ that probed by
Moorwood et al. (2000), which is the most similar previous survey in
redshift coverage and depth to
this one. Here we present the results of our pilot survey, and use the H$\alpha$ LF to
evaluate the H$\alpha$ luminosity density at $z=2.23$, and hence constrain
the evolution of $\rho_{\rm SFR}$. In \S2 we outline the details of the
survey and reduction techniques, \S3 describes the narrow-band
selection procedure, and in \S4 we present the results, 
discuss some of the individual properties and calculate the luminosity
function for the H$\alpha$ sample. 
We compare our results with previous estimates for this
epoch, and with the latest semi-analytic predictions for the H$\alpha$
luminosity density at this epoch. The integrated luminosity function
provides us with the $\rho_{\rm SFR}$, and we compare this to other measurements to
place our result in a cosmological context. We also present the first clustering analysis
of H$\alpha$ emitters at high redshift. 
The two-point angular correlation function yields the
real-space correlation length, which we compare to other populations
over cosmic time to gain insight into the likely descendants of these
galaxies. We summarise the work in \S5 and give a brief overview
of our future extension to the survey. 
Throughout we have assumed $h =0.7$ in units of
100\,km\,s$^{-1}$\,Mpc$^{-1}$ and
$(\Omega_m,\Omega_\Lambda)=(0.3,0.7)$. Unless otherwise stated, all
magnitudes are on the Vega scale.

\section{Observations and data reduction}

\subsection{Observations}

The COSMOS field was observed during 2006 May, November and December
with WFCAM on UKIRT, 
with the broad-band ($K$-band) and narrow-band (H$_2$S1, $\lambda_c =
2.121\mu$m, $\delta\lambda = 0.021\mu$m) filters. WFCAM's standard
`paw-print' configuration of four 2048$\times$2048 0.4$''$/pixel detectors
offset by $\sim$20$'$ can be macro-stepped four times to mosaic a contiguous
region of $\sim$$55'\times55'$ (Casali et al. 2007). 
In this way we mapped 3 paw-prints, or $\sim$0.62\,square\,degrees of
COSMOS (centred on 10$^{\rm h}$00$^{\rm m}$28.6$^{\rm s}$, +02$^{\rm
  d}$12$^{\rm '}$21.0$^{\rm ''}$ J2000). Our
narrow-band exposure time is $\sim$13.5\,ks/pixel (in order to reach
the equivalent continuum limit, the broad-band
$K$ exposures were $\sim$0.5\,ks/pixel). The seeing
was consistently of the order $\lesssim$1$''$. To help with Cosmic ray rejection
over the relatively long narrow-band exposures (40 or 60\,s), we used
the NDR (Non Destructive Read) 
mode, whereas CDS (Correlated Double Sampling) mode was used for the
broad-band exposures. 
The near-infrared observations were obtained in the standard way, following a
14-point jitter pattern. In addition to the jitter pattern, to improve
sampling of the PSF with the 0.4$''$ pixels of WFCAM, 
the narrow-band frames were microstepped in a $2\times2$ grid with
$1.2''$ offsets at each position. We summarise the observations in
Table 1.

\begin{table*}
\caption{Observation log of the COSMOS field. Coordinates are in J2000
  and refer to the centre pointing of a WFCAM `paw-print'. The average
seeing in the broad and narrow-band frames (stacked frames) is
$\sim$1$''$.}
\centering
\begin{tabular}{lcccccc}
\hline
Field I.D. & R.A. & Dec. & Dates observed & H$_2$S1 exposure time &
Continuum ($K$) depth \cr
 & (h\,m\,s) & ($\circ$ $'$ $''$) & &  (ksec.) & (3$\sigma$ mag) \cr
\hline
COSMOS 2 & 10~~00~~52 & +02~~10~~30 & 20--26 May / 5, 14--20 December 2006 & 15.9 & 20.1\cr
COSMOS 3 & 10~~00~~00 & +02~~23~~44 & 20--26 May / 5, 14--20 December 2006 & 13.4 & 20.0 \cr
COSMOS 4 & 10~~00~~53 & +02~~23~~44 & 13--17 November 2006& 13.6  & 20.1\cr
\hline
\end{tabular}
\end{table*}

\begin{figure}
\centerline{\includegraphics[width=3.5in]{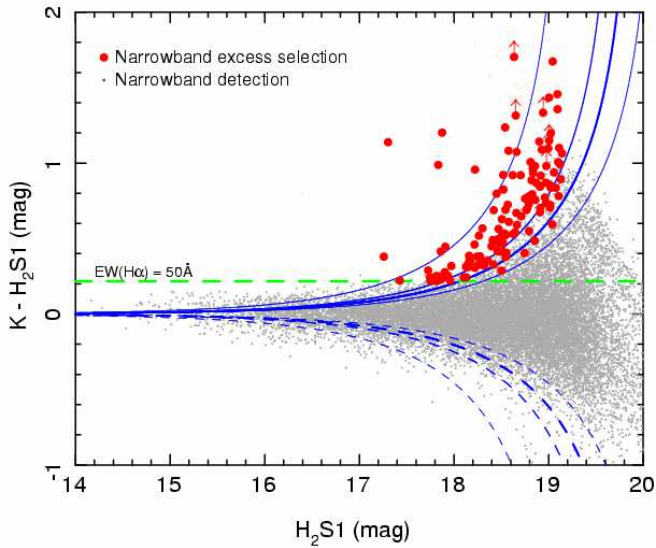}}
\caption[The $K-{\rm H_2S1}$ colour-magnitude diagram used to select
  narrow-band excess objects]{The $K-{\rm H_2S1}$ colour-magnitude diagram used to select
  narrow-band excess objects, and thus possible $z=2.23$ H$\alpha$
  emitters. Narrow-band colours have been corrected to account for
  broad-band continuum slope using the $(z-K)$ broad-band colour (see
  \S3.1). 
  $\Sigma=2,2.5,3~{\rm \&}~5$ selection cuts are shown, as is
  a 50\AA\ minimum in equivalent width used to select significant emitters. The
  $\Sigma$ cuts here are averaged over the whole field
  area, but these change slightly for each WFCAM tile as we take into
  account individual frame zeropoints and noise properties (although
  there is only a small variation over the survey). We show all
  detections that have would been selected as narrow-band excess sources
  as filled circles (limits indicate that source was not detected in the
  broad-band frame). These narrow-band excess sources are subject to a
  further selection described in more detail in \S3.2.}
\end{figure}

\begin{figure}
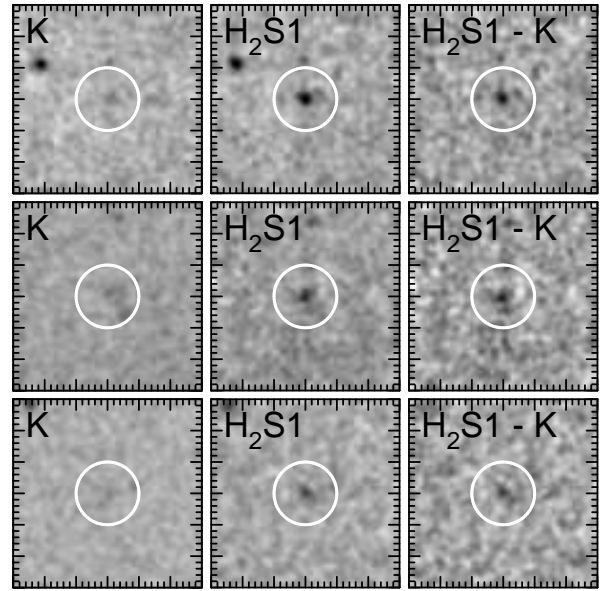

\begin{center}
\includegraphics[width=3.25in]{f2a.ps}\\
\includegraphics[width=3.25in]{f2b.ps}\\
\includegraphics[width=3.25in]{f2c.ps}
\end{center}
\caption[Examples of narrow-band excess objects]{Three examples
  of narrow-band excess sources of
  various significance, $\Sigma$. The top, middle and lower panels show
  $\Sigma\sim5$, $\Sigma\sim3$, $\Sigma\sim2.5$ respectively. The left
  hand panels show the $K$-band continuum, the centre shows the
  H$_2$S1 narrow-band image, and the right shows the
  continuum-subtracted narrow-band image. The $K$-band image has been
  scaled to account for the filter
  widths and both frames were locally registered with improved astrometric
  accuracy and smoothed with a Gaussian kernel to the same effective
  seeing. Images are $30''\times30''$ square and orientated North up and
  East left and scaled identically. Note that all images have been
  rebinned to the scale of the $K$-band frame: $0.4''$/pix. }
\end{figure}

\subsection{WFCAM data reduction}
For a given frame in a jitter
sequence of 14 images, we self-flat the data using a 
normalised median combination of the remaining frames in the sequence, taking care to 
mask-out bright objects in each frame. This is achieved by creating a
two-pass flat: in the first pass a rough flat field image is made from a
straight median combination of the data frames in order to reveal the
bright objects. In the next
step, we  use {\sc SExtractor}'s {\tt check\_image} output (v2.3.2 Bertin \&
Arnouts 1996) to identify regions where bright stars and galaxies lie
and then mask these when constructing the master flat. This
minimises the 
artifacts seen around bright sources. Some care must be taken
with the micro-stepped frames -- the small offset between consecutive
frames in a micro-sequence are smaller than many extended
sources, therefore we construct flats for each position in the  micro-stepped sequence
separately as if they are part of a standard dither.
  
A world coordinate system is fit to each frame by querying the
USNO\,A2.0 catalogue, fitting on average $\sim$100 objects, 
and then frames are aligned and
co-added with {\sc swarp}\footnote{Both {\sc swarp} and {\sc
    SExtractor} are distributed by Terapix ({\tt
    http://terapix.iap.fr}) at the Institut d'Astrophysique de Paris.}
(this performs a background mesh-based sky subtraction, optimised to
provide the best representation of the large scale variations across
the field). In the case of the micro-stepped (narrow-band images), this step is preceded by
drizzling the frame onto a grid with a pixel scale of 0.2$''$
(i.e. half the camera pixel scale). 

WFCAM frames suffer from significant cross-talk, manifesting itself
in toroidal features at regular (integer 
multiples of 128) pixel intervals from sources in the read-out direction. 
As these are tied to the positions of real objects, they are not
removed by dithering the frames. High order cross-talk can easily
mimic real objects, therefore these positions are flagged at the
catalogue stage; we discuss this further in \S3.

Broad and narrow-band frames are calibrated by matching $10 < K <
15$ stars from the 2MASS All-Sky Catalogue of Point Sources (Cutri et al. 2003) which are
unsaturated in our frames. Since the narrow-band filter falls in the
$K$-band, we can check the zero-point offsets to be applied to the narrow-band
frames directly from the broad-band calibration, taking into
account the relative widths of the filters, such that the offset
between the zeropoints is $2.5\log(\Delta\lambda_K /
\Delta\lambda_{\rm H_2S1})$. As an additional verification, we check that $K-{\rm
  H_2S1} \sim 0$ as expected using the photometry of the bright
stars. We estimate our absolute $K$ calibration is good to $\lesssim$1\% from
the scatter compared to 2MASS\footnote{An independent
  data reduction of a sub-set of the science frames by the Cambridge
  Astronomical Survey Unit (CASU) confirms consistent photometry (M. Riello
  private communication).} 

\subsection{Source extraction and survey limits}

Our survey is made up of a mosaic of three WFCAM paw-prints, i.e. twelve
$13.7'\times 13.7'$ tiles.
Sources are detected and extracted using {\sc SExtractor} (Bertin \&
Arnouts 1996).  
A final narrow-band catalogue is made by matching detections in the H$_2$S1 to the broad-band
image. Since the frames are well aligned  astrometrically, a
simple geometric matching algorithm suffices, with a maximum search
radius of 2$''$. Cross-talk artifacts in WFCAM frames can mimic
the narrow-band excess objects we are selecting for. Therefore, to
clean the catalogue from possible cross-talk contaminants we measure
the chip $(x,y)$ position of bright sources, and
flag regions at integer multiples of 128 pixels along the
direction of readout. These chip positions can be converted to a sky
coordinate and then compared to the narrow-band catalogue. As a rule, we
remove an object from the catalogue if it lies within 2$''$ of any
cross-talk position. Note that by avoiding the zeroth order cross-talk
position, we also avoid spurious detections close to bright stars
(although in this case the exclusion radius is larger to account for
extended halos and diffraction spikes, at 30$''$). Due
to the lower exposures received by the edges of the individual frames
(caused by the jitter pattern), we ignore detections from within
$\sim$10$''$ of any frame edge. The total area lost due to cross-talk
and star masking is negligible ($<0.1$\%), however including the
edge-clipping we survey an effective area of 0.603\,square degrees.  

 Due to the fact that the observations were
acquired over a long period of time, with slightly different seeing
and exposure times (Table 1), we treat each tile separately, detecting objects
down to each tile's limit. The average 3$\sigma$ depth of the
broad-band frames is $K=20.1$\,mag; we detect a total of 19079 objects
down to this limit across 0.603\,sq.~degrees. 
Note that all magnitudes are measured in 3$''$
diameter apertures, and this recovers the majority of the flux. 

To confirm the depth and completeness of the images, we perform Monte
Carlo simulations of the detection of a large number of `fake'
sources (a single Gaussian point source generated with {\sc iraf}'s {\sc
  artdata} with a FWHM matching the image PSF). 
By scaling the artificial source over a range of known flux,
and inserting at random locations over the frames, we measure the detection
rate (i.e. the completeness) as a function of observed magnitude using an
identical extraction procedure to that used to generate the main
catalogue. We use this detection completeness information to correct
the derived luminosity function of H$\alpha$ emitters.

\begin{figure*}
\centerline{\includegraphics[width=0.9\textwidth]{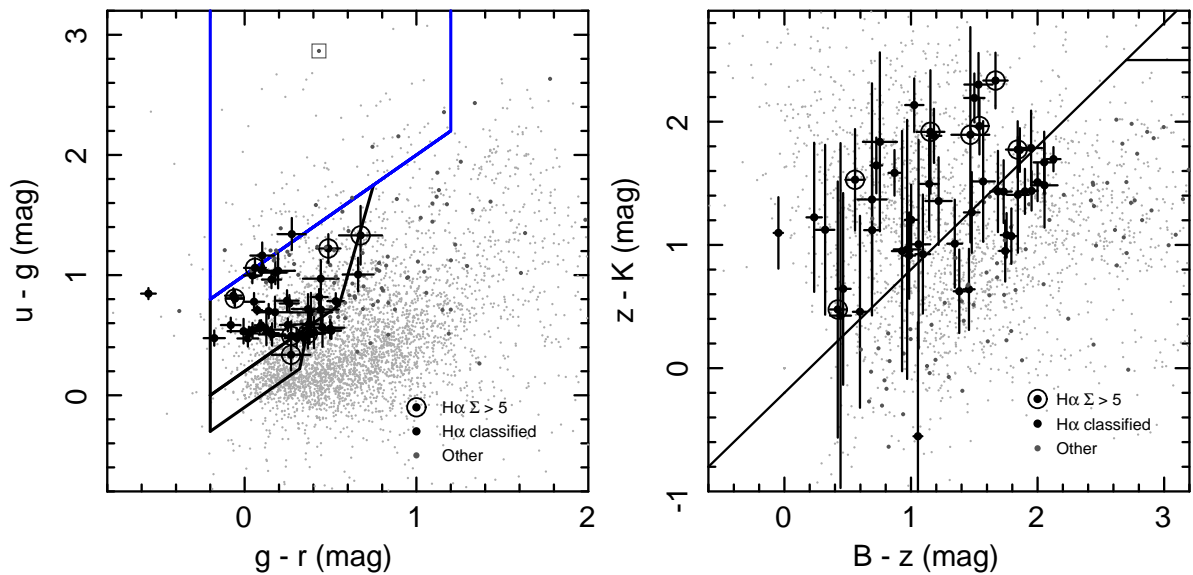}}
\caption[Broad-Band {\it ugr} and {\it BzK} colours of narrow-band
excess objects]{Broad-band colours of narrow-band selected
  objects from our survey. (left) {\it ugr} colours with the LBG and BM/BX selection criteria of Steidel
et al. (2003) (right) {\it BzK} colours divided into high redshift
starforming galaxies ({\it sBzK}) and high redshift passive galaxies
({\it pBzK}) and foreground galaxies, using the selection of Daddi et
al. (2004). In both cases we compare to a 1-in-100 sampling of the  general colour distribution using
photometry available for the COSMOS field.
It is clear that the straightforward narrow-band selection
outlined in \S3.1 is also detecting low redshift galaxies, and we
require further selection with broad-band colours to identify
these. However, the low redshift detections likely correspond to
{\it real} emission lines, and we discuss this further in \S3.2. Large
points indicate our secure H$\alpha$ emitters after employing our
secondary colour selection outlined in \S3.2 and we highlight the high
significance ($\Sigma>5$) candidates, noting that all of these fall in
the {\it sBzK} and BX/BM selection regions. For clarity, we only show
photometric uncertainties for H$\alpha$ candidates. }
\end{figure*}

\section{Selection technique}

\subsection{Narrow-Band excess selection}

Potential $z=2.23$ H$\alpha$ emitting candidates are initially
selected based on the significance of
their narrow-band excess, i.e. their $K - {\rm H_2S1}$ colour. Galaxies
with a strong emission line falling in the narrow-band filter will have 
${\rm H_2S1} < K$. The significance of the narrow-band excess can be
parameterised in terms of $\Sigma$, which quantifies the excess
compared to what would be expected for a source with a flat spectrum 
(Bunker et al. 1995).
This parameter is similar to a standard signal-to-noise 
selection, and we chose to select candidate emitters with $\Sigma >
2.5$. This is equivalent to a colour of:
\begin{equation}
m_{\rm K} - m_{\rm H_2S1} = -2.5\log\left[1-\Sigma \delta 10^{-0.4(c-m_{\rm H_2S1})}\right]
\end{equation}
where $c$ is the zeropoint of the narrow-band image and $\delta$ is the
photometric uncertainty. The H$\alpha$ line flux $F_{\rm H\alpha}$ and
equivalent width $W_{\rm  H\alpha}$ can be calculated:
\begin{equation}
F_{\rm H\alpha} = \Delta\lambda_{\rm H_2S1}  \frac{f_{\rm H_2S1} - f_K}{1-(\Delta\lambda_{\rm H_2S1} /\Delta\lambda_K) }
\end{equation}
and
\begin{equation}
W_{\rm H\alpha} = \Delta\lambda_{\rm H_2S1}  \frac{f_{\rm H_2S1} -
  f_K}{f_K-f_{\rm H_2S1}(\Delta\lambda_{\rm H_2S1} /\Delta\lambda_K) }
\end{equation}
where $\Delta\lambda_K$, $\Delta\lambda_{\rm H_2S1}$ are the widths of
the broad and narrow-band filters and $f_{K}$ and $f_{\rm H_2S1}$ are
the flux densities in the two filters. 
$W_{\rm H\alpha}$ is simply the ratio of the line flux and continuum
flux density, and will also include contribution from the adjacent
{\sc [Nii]} line. 
However, we calculate the expected H$\alpha$ line flux by assuming $F_{\rm [NII]} /
F_{\rm H\alpha} = 0.33$ for extragalactic H{\sc ii} regions
(Kennicutt \& Kent 1983)  and correcting appropriately. Note that this
may result in conservative H$\alpha$ fluxes, since for metal-poor
galaxies this contribution may be lower.

In Figure~1 we present the colour-magnitude diagram with
selection criteria. 
Some care must be taken in deciding on the
selection, and we briefly describe it here. Firstly, we only use
narrow-band detections with detection
significances of $>3\sigma$. It is not necessary that the source is
also detected in
the broad-band frame, and we assign an upper limit if there is no
corresponding continuum detection. 
We define an initial selection criteria of $\Sigma>2.5$ (this is a
function of narrow-band flux and combined error due to sky noise in
both the broad and narrow-band images, as described above). 
We  account for bright foreground objects with
steep continua across the $K$-band by making use of the $z-K$ colours
available for all detections from the COSMOS archival $z$-band
imaging (Capak et al. 2007). Their very red continua will result in a
larger effective wavelength for the broad-band detections. We  correct
for this  to make a better estimate of the continuum contribution to the narrow-band
flux at the 2.121$\mu$m. We measure $K-{\rm H_2S1}$ as a
function of $(z-K)$, and fit the data with a linear trend using a
least squares fit. The slope of this line is then be used to
correct the $K-{\rm H_2S1}$ colour. We find that on average the correction
factor to the narrow-band magnitude excess is $0.03(z-K)$. To account
for narrow-band detections  that have large
significance but low equivalent widths (e.g. van der Werf
et al.\ 2000),  we enforce an additional selection criteria of $W_{\rm
  H\alpha} > 50$\AA.
This is an arbitrary threshold, but is chosen to reflect the general
scatter about  $K-{\rm H_2S1}=0$ (see Figure 1). 

The H$\alpha$ line-flux limit is the theoretical minimum that could be
detected taking into account all these selection criteria. The average
$3\sigma$ H$\alpha$ flux limit over the survey is
$1\times10^{-16}$\,erg\,s$^{-1}$\,cm$^{-2}$.
As a final quality control check, we visually inspect each candidate
in order to filter any remaining erroneous sources (for example,
non-crosstalk associated artifacts, etc.). In Figure 2 we present the
narrow-band field around three candidate emitters representing a range of
detection significance ($2.5 \lesssim \Sigma \lesssim 5$). For each panel
we show the narrow-band frame before and after continuum
subtraction using the broad-band image after suitably scaling to
account for the relative filter widths. We detect a total of 180
sources meeting the narrow-band selection criteria across 0.603\,square
degrees.

\subsection{Broad band colours: further selection, redshift
  confirmation and line contamination estimates}

We must be certain that the
selection in colour space is actually identifying the H$\alpha$ line at
$z=2.23$ rather than another strong emission line at another redshift. 
At higher redshift, the most important contaminant is {\sc
  [Oiii]}$\lambda5007$ at $z=3.23$
 (although these objects are interesting in their own right as
 possible examples of high-$z$ AGN or star-forming galaxies). At low
 redshift, we are potentially susceptible to any strong line with an
 emission wavelength between
 H$\alpha$ and H$_2$S1. 

The most robust way to confirm the selection is through
follow-up spectroscopy which can unambiguously identify the
lines. However, complete spectroscopic follow up (at
least in the short-term) is unfeasible. A `cheaper' alternative is to
make use of extensive multi-band optical/near-infrared photometry
available for COSMOS (Capak et al.\ 2007), which can be used to
improve the H$\alpha$ selection. Unfortunately, at $z=2.23$ the current photometric redshifts are
not reliable enough to confirm H$\alpha$ emission (Mobasher et
al. 2007),  so we turn to the cruder method of colour/colour-selection
to clean our high redshift sample.

We therefore define a secondary optical selection, tuned according to the
colours of a sub-sample of high significance ($\Sigma > 5$) H$\alpha$
candidates from the pure narrow-band selection.
To discriminate between high and low redshift emission line galaxies,
we first use the {\it sBzK} selection first presented by Daddi et al.\
(2004), and now becoming popular to select active galaxies at $z>1.4$.
An alternative selection of high redshift galaxies is made with
{\it ugr} colours, and has been used to search for $z\sim3$
Lyman Break galaxies (LBGs, Steidel et al. 1996) and  `BM/BX' 
star-forming galaxies at $z\sim2$ (Adelberger et al. 2004, Steidel et
al.\ 2003). We use the BM/BX selection as a further indicator
that the narrow-band selected galaxies are at high redshift
(additionally the LBG selection can also be used to detect possible
$z\sim3$ [O {\sc iii}]$\lambda 5007$ emitters). Candidate H$\alpha$
emitters are
narrow-band excess objects which also satisfy the BM/BX
selection {\it or} the {\it sBzK} selection. In a further effort  to eliminate
foreground contaminants we impose a magnitude cut of $z < 22.5$
(equivalent to $M_u < -23.7$ at $z=2.23$). This
yields a total H$\alpha$ sample of 55 compared to our original selection of
180 sources on the basis of narrow-band selection alone. 
We present the colour-colour plots with selection criteria in
Figure~3. After the colour selection, the majority of H$\alpha$ emitters
fall in both the BM/BX and {\it sBzK} regions, with those that do not
having large enough photometric errors that place them $<1\sigma$ away
in colour-space. Note that two objects flagged as H$\alpha$ emitters
are potentially low-redshift [Fe{\sc ii}]$\lambda 1.64$ emitters based
on their photometric redshifts (see Figure 4 and discussion
below). Thus, the photometric selection is not a fool-proof method,
although it suggests the contamination rate is low ($\lesssim3$\%). To
estimate the number of galaxies that could have been potentially
missed by this further colour selection, we Monte Carlo our selection
by allowing the photometry to vary
randomly in a range defined by the (1$\sigma$) uncertainty on each point, then count
the number of galaxies satisfying the H$\alpha$ criteria  and compare
this to  the nominal value. Repeating this procedure many 
times builds up a measure of the the likely
contamination/incompleteness as galaxies are scattered in and out of
the selection regions. Our results suggest that the number of galaxies
missed (or alternatively the interloper fraction) could be as much as
10\%. Assessing this contamination more accurately will require follow-up
spectroscopy, which at the time of writing, is underway.  
   
What are the remaining narrow-band excess objects which fail our colour
selection? At least one appears to be a high
redshift (possibly [O{\sc iii}]$\lambda 5007$) emitter selected in the LBG cut, but
the bulk of the remaining objects are likely to be low-redshift given
their colours. There are several other low redshift
emission lines that could potentially be detected, and
to investigate this, in Figure~4 we plot the probability distribution
$P(z)$ of $z_{\rm phot}$ of (a) candidate H$\alpha$ emitters, (b) Pa$\alpha$ and
Pa$\beta$ candidates and (c) unclassified remaining narrow-band excess
detections.  Using a conservative photometric redshift uncertainty of
$dz=\pm0.05$ we identify 12 Pa$\alpha$ and 6 Pa$\beta$ candidates.
In the unclassified redshift distribution there are two clear peaks in
the distribution at $z\sim0.4$
and $z\sim0.9$. The low redshift emitters could potentially be [Fe{\sc
  ii}]$\lambda$1.6$\mu$m or several Brackett lines, whereas the
slightly higher redshift peak
might correspond to Pa$\gamma$. Clearly this must be confirmed through
spectroscopy, but it appears that the narrow-band selection is
detecting real line-emitters, but is not sufficient alone to isolate
H$\alpha$ at $z=2.23$.  Identification of lower redshift interlopers
will become easier when more accurate photometric redshifts are available
for COSMOS (Mobasher et al. 2007 \& private communication).

To check for contamination for higher-$z$ line emitters, we use a {\it
  ugr} cut to select $z\sim3$ Lyman Break
  galaxies (LBGs), and so identify possible [O{\sc
    iii}]$\lambda 5007$ contaminants (Figure 3). In our total narrow-band selected sample, only
  one object satisfies the LBG selection criteria, and therefore
  we conclude that the $z=3.23$ [O{\sc iii}]$\lambda 5007$ contamination rate is
  negligible for our sample. We note that in a follow up spectroscopic
  survey of narrow-band identified $z=2.2$ H$\alpha$ emitters by Moorwood et
  al. (2000), six of 10 candidates were confirmed as $z=3.23$ [O{\sc
    iii}]$\lambda 5007$. Thus it is not clear whether environmental effects are
  responsible for the high contamination observed in this previous
  small scale survey. 

Finally we assess the impact of contamination from AGN at $z=2.23$
which would impact on our assessment of $\rho_{\rm SFR}$ from the H$\alpha$ LF. 
We are probing the expected peak in the SFRD, but this is also
associated with the peak of AGN activity (Shaver et al.\ 1996; Boyle
\& Terlevich 1998) -- what is the expected contamination rate in our
H$\alpha$ selected sample? As with contamination from
emission lines at different redshifts, without spectroscopic
information (for example the H$\alpha$/{[N{\sc ii}] ratio
  discriminating between star-formation and nuclear activity), we must
  address this issue statistically. As described by van der Werf et
  al.\ (2000), the AGN interloper fraction is likely to be low: the fraction
  of AGN in local H$\alpha$ selected galaxies is $<$5\% (Gallego et
  al.\ 1995), and this is
  not likely to increase toward high redshift (Teplitz et al.\
  1998).  On the other hand, Shioya et al. (2008) include a 15\%
  correction for AGN contamination in the H$\alpha$ luminosity
  function at $z=0.24$, based on the results of Hao et al.\ (2005) who
  studied the H$\alpha$ luminosity density from AGN in the Sloan
  Digital Sky Survey. We adopt this same 15\% AGN contamination in
  our measurement, resulting in a more
  conservative estimate of $\rho_{\rm SFR}$.

\begin{figure}
\centerline{\includegraphics[width=3.25in]{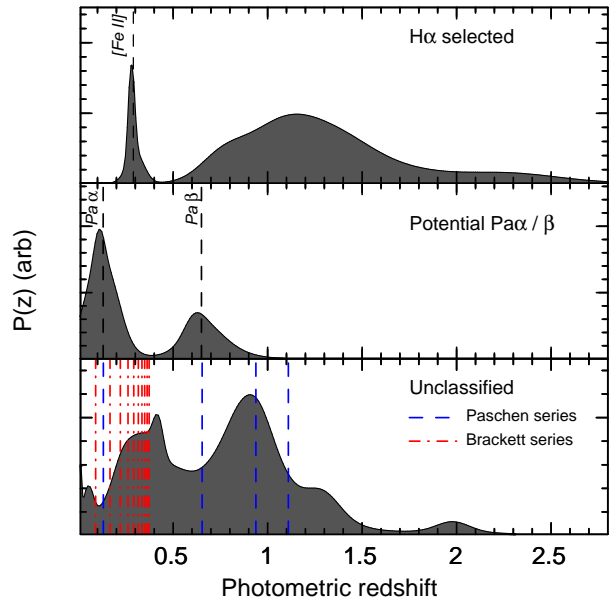}}
\caption[Photometric redshift distribution of narrow-band excess
detections]{The summed  probability distributions $P(z)$ for photometric
  redshifts: (top) narrow-band excess detections satisfying our
  {\it ugr}/{\it BzK} selection. Note that there is a spike in
  $P(z)$ at $z\sim0.3$, corresponding to the redshift of [Fe{\sc
    ii}]$\lambda 1.64$
  in the H$_2$S1 filter. This spike represents two objects and these are
  removed from our sample. The broad distribution of $z_{\rm phot}$ at
  $z\gtrsim 1$ are probably real H$\alpha$ emitters with poorly determined
  photometric redshifts; (middle) distribution of galaxies selected as
  Pa$\alpha$ and Pa$\beta$ emission lines with $z_{\rm
    phot}=0.13\pm0.05$ and $z_{\rm phot}=0.65\pm0.05$ respectively;
  (bottom) remaining  `unclassified' emitters. There are two main
  peaks to the distribution of unclassified objects: at $z\sim0.3$
  which could correspond to [Fe{\sc ii}]$\lambda 1.64$ or several Brackett lines. At
  $z\sim0.8$ the peak could correspond to Pa$\gamma$. 
}
\end{figure}

\subsection{Spectroscopic comparison}

The {\it z}COSMOS survey (Lilly et al.\ 2007) 
is an extensive spectroscopic project
obtaining thousands of redshifts in the COSMOS field. To search for any existing
spectra corresponding to our narrow-band excess sample, we
cross-correlate our catalogue with the current {\it z}COSMOS
spectroscopic catalogue (S. Lilly priv. comm.). We do not expect a large
degree of overlap between our catalogue and {\it z}COSMOS, since we
are detecting relatively faint galaxies, and indeed there
were only five matches with secure redshifts. One
H$\alpha$ candidate is included in the `faint' {\it z}COSMOS catalogue and
confirmed to securely lie at $z=2.2329$, verifying the selection at
least for this candidate. 
The remaining matches are all in the `bright' sample. Three of these
were identified by us as being possible Pa$\alpha$ at $z\sim0.13$, and
indeed two have confirmed redshifts of $z=0.1312$
and $z=0.1298$, although another has $z=0.1086$. The final match has
$z=0.7340$, and it is unclear if this corresponds to a real emission line. 
Although this is a very limited
spectroscopic confirmation, it does suggest that we are detecting a
range of emission lines, and that our secondary filtering technique is
useful for weeding out low redshift interlopers. We note that our near-infrared
spectroscopic follow-up will allow us to make a much better
estimation of the contamination rate.

\section{Results}

We detect a total of 55 robust candidate $z=2.23$ H$\alpha$ emitters over
0.603\,square degrees of the COSMOS field, resulting in a surface
density of $91\pm10$\,deg$^{-2}$ above a flux limit of
$1\times10^{-16}$\,erg\,s$^{-1}$, equivalent to a SFR of
30\,$M_\odot$\,yr$^{-1}$ at this redshift.
 In the following sections we briefly discuss their 
 multi-wavelength and morphological properties, and use the sample to evaluate the
H$\alpha$ luminosity function and thus star-formation rate density at
$z=2.23$. Finally, we make an estimate of the real space correlation
length of H$\alpha$ emitters by measuring their two-point angular
correlation function.

\subsection{H$\alpha$ luminosity density at $z=2.23$}

\subsubsection{Evaluating the luminosity function of H$\alpha$ emitters}

To calculate the luminosity function of H$\alpha$ emitters, we convert
line fluxes to luminosities, correcting for 33\% [N{\sc ii}]
contribution to the measured flux. The volume is reasonably
well defined by the survey area and narrow-band filter width (50\%
cut-on/off: 2.111--2.132$\mu$m), which probes a
co-moving depth of $\sim$40\,Mpc, and is equivalent to
220$\times$10$^3$\,Mpc$^3$ (co-moving). However the reader should note
that since the narrow-band transmission function is not a perfect
top-hat, the effective volume probed will vary as a function of intrinsic
luminosity, with more luminous H$\alpha$ emitters being detected over a larger volume than
fainter sources. Similarly, the incompleteness will vary strongly as a
function of apparent luminosity, since faint H$\alpha$ emitters at redshifts
corresponding to the low-transmission wings of the narrow-band filter
might be lost, and bright lines will have their luminosity
underestimated, contributing to the counts in lower luminosity
bins. To estimate the impact of this effect on our measured results,
we model these observational effects on a `true' model luminosity
function. 
Our simulation consists of a set of model H$\alpha$
emission lines convolved with the H$_2$S1 transmission function, where
the lines are randomly placed in a redshift range corresponding to the
full width of the filter. After convolving 1000 such lines per
luminosity interval, we recover the effective
volume probed, the incompleteness and luminosity biases (i.e. including
intrinsically bright emitters dimmed by the edge of the transmission
function falling in lower luminosity bins). The combination of these effects
modifies the model luminosity function, the main change being a
$\sim$10\% reduction in the faint-end slope $\alpha$ and a $\sim$20\%
decrease in $\phi^\star$ compared to a uniform transmission model. The effect on
$L^\star$ is negligible. As we describe below, since we fix the faint end
slope $\alpha=-1.35$, the difference in the integrated luminosity
function above the simulated detection threshold is slightly less than
if this was a free parameter, $\sim$10\%, and we
correct for this difference in our measurement of $\rho_{\rm SFR}$.
 
Although H$\alpha$ is relatively insensitive to dust extinction
(unlike Ly$\alpha$, it is not resonantly scattered), dust absorption
must still be taken into consideration when deriving luminosities. A robust
technique would be to obtain spectroscopy of each source to measure
the Balmer H$\alpha$/H$\beta$ decrement, but this is
not currently  feasible for our sample. Instead we simply adopt $A_{\rm
  H\alpha}=1$\,mag (Kennicutt 1992; 1998), as is applied in similar
earlier studies (Fujita et al.\ 2003; Pascual et
al.\ 2005; Ly et al.\ 2007) for SFR estimates. 
Although it is known that the specific
extinction is a function of SFR (Jansen et al.\ 2001;
Arag\'{o}n-Salamanca et al.\ 2003; Hopkins et al.\ 2001), for ease of
comparison with other samples, we apply
this constant correction to our sample, noting that in the cases where
the actual extinction is higher, our measured SFRs will be
conservative. We compare SFR indicators for our sources in \S4.2.2 and
confirm that the correction used here is appropriate.

Our LF has been corrected for completeness on a bin by
bin basis using the completeness functions estimated from the
narrow-band simulation (\S2.3), and the lower luminosity limits have
been dealt with by performing a survival analysis on the binned data,
using the probability estimator of Avni et al. (1980). Note that we
apply a correction for the incompleteness caused by the shape of the
H$_2$S1 transmission described above on the integrated luminosity
function (see \S4.1.2). The errors
are derived from the Poisson statistics of each bin, although we apply a
slightly more conservative bootstrap analysis to estimate the
uncertainties on the fit values, described in more detail below.
We fit the LF with a Schechter function characterised
by $\alpha$, $\phi^\star$ and $L^\star$: 
$\phi(L){\rm d}L = \phi^\star(L/L^\star)^\alpha \exp(-L/L^*){\rm d}(L/L^\star)$.
Since we lack sufficient depth to fit the faint end slope, we fix
$\alpha=-1.35$,  the value observed by  Gallego et al. (1995)
for their $z=0$ H$\alpha$ study. This approach was also used by Yan et al. (1999) for
the H$\alpha$ LF at $z=1.3$. Therefore we cannot show any
evolution in the faint end slope of the H$\alpha$ LF -- a deeper
survey is required for this. In fact Reddy et al. (2007) find evidence for a
steeper faint end slope for UV/optical colour-selected star-forming galaxies at
high redshifts ($z\sim2$--$3$). Adopting a $\alpha < -1.35$ would
obviously have a rather dramatic effect on extrapolated values for $\rho_{\rm
  SFR}$, which we discuss in \S4.1.2. 

To estimate uncertainties on the LF, we
refit the Schechter function by bootstrapping, with the variance of
the resulting parameters forming our uncertainty in $\phi^\star$ and
$L^\star$. We find $\phi^\star=(1.45\pm0.47)\times10^{-3}$\,Mpc$^{-3}$
and $\log L^\star = {42.83\pm0.13}$\,erg\,s$^{-1}$.  
We present the results in Figure~5 in comparison
to lower redshift studies of the H$\alpha$ LF $z=0$, demonstrating the
sharp increase in H$\alpha$ emitters from $z=0$--2.23. 
Our results are very similar to the luminosity distribution
found by Yan et al. (1999), suggesting that there is little evolution of the
bright end of the H$\alpha$ LF across the 2\,Gyrs between $z=2.23$ and $z=1.3$.

\begin{figure}
\centerline{\includegraphics[width=3.4in]{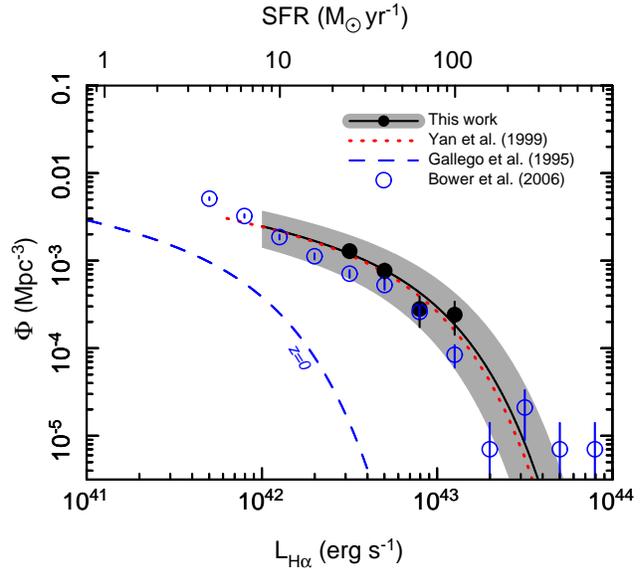}}
\caption[Luminosity function of H$\alpha$ emitters at $z=2.23$]{H$\alpha$
  luminosity function from this work compared to
  those derived at $z=0$ (Gallego et al. 1995) and $z=1.3$ (Yan et
  al. 1999) (these are not corrected for dust extinction). 
  Errors are based on the
  Poisson counting error and the completeness correction applied to
  each bin, derived from the detection rate of narrow-band selected
  sources. The luminosity function is fit with a Schechter function
  with a fixed faint end slope, $\alpha=-1.35$ (identical to Yan et
  al. 1999 and Gallego et al. 1995) -- deeper observations will be
  required to probe the evolution of the faint end slope of the
  luminosity function. Our results show significant evolution from
  $z=0$ equivalent to an order of magnitude in luminosity evolution.
Our LF is also very similar to the $z=1.3$
  H$\alpha$ luminosity function, consistent with little or no evolution
  between $z=2.23$ and $z=1.3$, interpreted in terms of the star
  formation rate density as a peak or plateau occurring at $z\sim2$
  (see Figure 6). We compare to the theoretical prediction of the
  luminosity distribution of H$\alpha$ emitters from the Bower et al.\
(2006) variant of the semi-analytic {\sc galform} code (Cole et
al. 2000). The observed counts agree well with the model, although
there are hints that the faint end slope is slightly steeper than our assumed
value.}
\end{figure}

\begin{figure*}
\centerline{\includegraphics[width=4in]{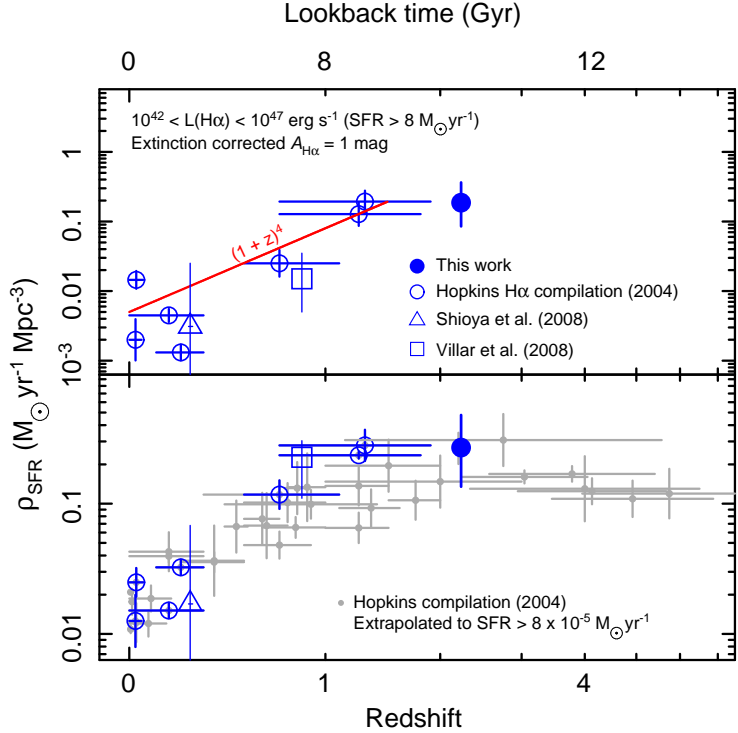}}
\caption[Evolution of the star-formation rate density out to
$z=2.23$]{The evolution of the star-formation
  rate density out to $z=2.23$ using the
  H$\alpha$ tracer alone, and compared to a heterogeneous mix of multi-wavelength
  tracers (all converted to the cosmological parameters used
  throughout this work). 
  The data are collected taken from the
compilation of literature data made by Hopkins (2004), and recent
works of Shioya et al. (2008) and Villar et al. (2008).
Where corrected for extinction, are modified by a common factor $A_{\rm
  H\alpha}=1$\,mag. In the top
panel, we calculate the integrated H$\alpha$ luminosity down to the
approximate luminosity limit of the present survey, $L_{\rm H\alpha} >
10^{42}$\,erg\,s$^{-1}$. Our results support a flattening of the $\rho_{\rm SFR}$
between $z$$\sim$1--2.
In the bottom panel we apply the same
integration range as Hopkins (2004), extrapolating down to $L_{\rm H\alpha} >
10^{37}$\,erg\,s$^{-1}$. The general evolution of $\rho_{\rm SFR}$ is mirrored
here, but the reader should be aware of the potentially strong impact
uncertainties in the faint end slope have on the integrated luminosity
function. Surveys of high-redshift star-forming galaxies have yet to
satisfactorily constrain this parameter.}
\end{figure*}

\subsubsection{The star-formation rate density at $z=2.23$}

Next we use the integral of the LF to find the 
volume averaged star-formation rate density, $\rho_{\rm SFR}$. In
Figure~6 we present the evolution of $\rho_{\rm SFR}$ out to $z=2.23$ using
H$\alpha$ {\it only}. In order to compare results, we take the LF
parameters for H$\alpha$ surveys compiled by Hopkins (2004) and
integrate these down to $L_{\rm H\alpha} = 10^{42}$ \,erg\,s$^{-1}$,
the approximate luminosity limit of the present survey.
To convert H$\alpha$ luminosity to a star-formation rate, after
applying a factor 2.5 to $L_{\rm H\alpha}$ to account for $A_{\rm H\alpha}=1$\,mag,
we use the standard calibration of  Kennicutt (1998); ${\rm SFR[H\alpha]
  } = 7.9\times10^{-42}(L_{\rm H\alpha} /{\rm
  erg\,s^{-1}})\,M_\odot\,{\rm yr^{-1}}$ (assuming continuous star
formation, Case B recombination at $T_e = 10^4$\,K and a Salpeter
initial mass function [IMF] ranging from
0.1--100$M_\odot$). Correcting for the transmission incompleteness
described in \S4.1.1 and a 15\% correction for AGN contamination
(Shioya et al. 2008),  we find
$\rho_{\rm SFR} =
0.17^{+0.16}_{-0.09}$\,$M_\odot$\,yr$^{-1}$\,Mpc$^{-3}$
where the uncertainty 
is equivalent to the 1$\sigma$ uncertainties on the luminosity
function fit parameters. Our results
support a flattening or peak in the level of star-formation activity
between $z$$\sim$1--2.

In Figure~6 we also show the evolution of $\rho_{\rm SFR}$, but extrapolated to
a much fainter luminosity, $L_{\rm H\alpha} > 10^{37}$\,erg\,s$^{-1}$ (${\rm SFR} >
8\times10^{-5}\,M_\odot\,{\rm yr^{-1}}$), in line with the
$\rho_{\rm SFR}$ compilation of Hopkins (2004). We remind the reader of the
uncertainty inherent in interpreting $\rho_{\rm SFR}$ estimates derived from such
extrapolations: the uncertainty in the slope of the faint end of the
luminosity function could have a dramatic effect on the result.
For this extrapolated case (again, corrected for extinction), we find $\rho_{\rm SFR} =
0.25^{+0.20}_{-0.13}$\,$M_\odot$\,yr$^{-1}$\,Mpc$^{-3}$. This is in broad
agreement with the range of $\rho_{\rm SFR}$ estimates (for a variety of tracers)
at $z=2.23$ (Fig.~6) and confirms the plateau in $\rho_{\rm SFR}$
above $z\sim1$.

Our measurements of $\rho_{\rm SFR}$ are consistent with the results of Reddy et
al. (2008), who estimate the  H$\alpha$ luminosity density at $1.9\leq
z<2.7$ by using the correlation between
dust-corrected ultra-violet and H$\alpha$ SFRs (Erb et al. 2006). They
find $\rho_{\rm SFR} = 0.35\pm 0.09$\,$M_\odot$\,yr$^{-1}$ over this
(albeit larger) redshift range. Reddy et al. (2008) measure a steeper faint end slope
for the ultraviolet luminosity density at $z\sim2$, $\alpha =
-1.88$. If this behaviour is reflected in the general star-forming
population at this redshift, then we would also expect a steeper
H$\alpha$ faint end slope. This would increase the extrapolated
$\rho_{\rm SFR}$ presented here. For example, fixing $\alpha=-1.65$
(more in line with the theoretical luminosity distribution from Bower
et al. 2006), $\rho_{\rm SFR}$ would increase by nearly a factor of
2. This should be taken as a cautionary point by the reader, when
interpreting our extrapolated results.

There is a final caveat regarding the
luminosity function presented in Figure 5; namely the treatment of
dust. In this analysis we have assumed a uniform correction, $A_{\rm
  H\alpha}=1$\,mag for ease of comparison with other results (we
discuss the validity of this assumption in \S4.2.2). Note
that, in reality, this correction may be a function of luminosity,
such that the most luminous systems require larger corrections. This
would have most effect on the bright end of the LF, where we suffer
the most from low-number statistics. 

  \begin{figure*}
\centerline{\includegraphics[width=\textwidth]{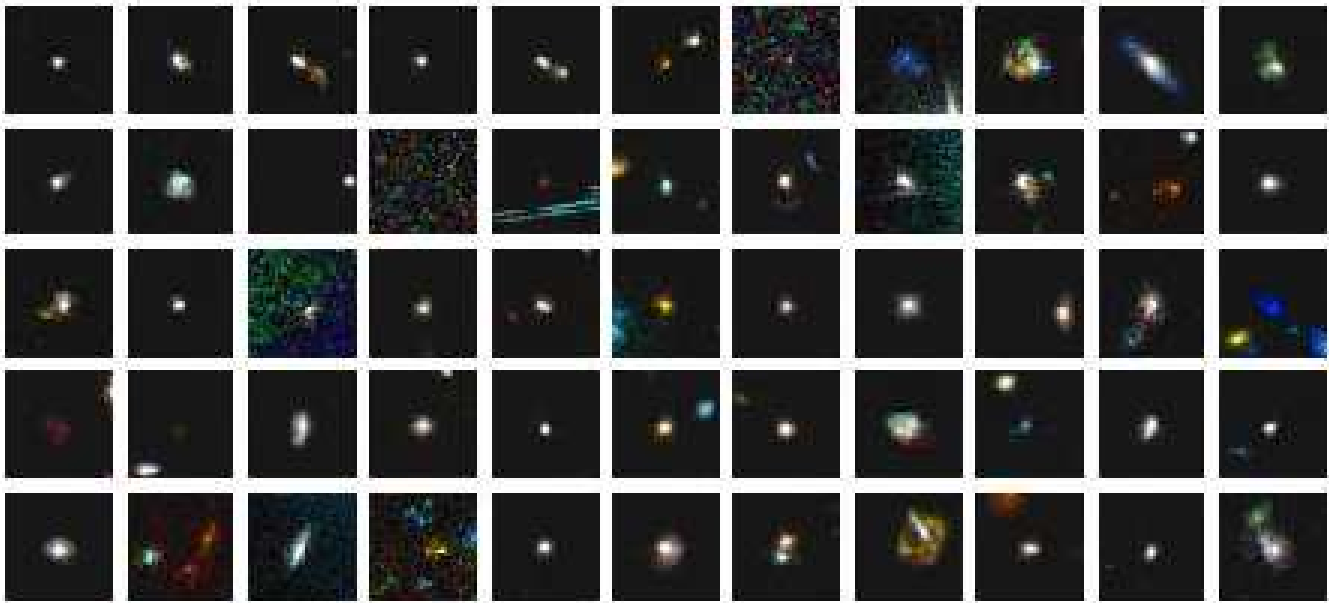}}
\caption[{\it ACS} thumbnail images of 55 H$\alpha$ candidates]{
  $5''\times5''$ {\it HST} ACS thumbnail images of the 55
 $z=2.23$ H$\alpha$ candidates. The images are coloured with a map
 generated from the ground based {\it Brz} imaging, and use the F814W
 filter as a luminance channel (all images are scaled
  identically and orientated with north up and east to the left). Note
  that at $z=2.23$, $5''$ is equivalent to $\sim$40\,kpc in
  projection. The images are organised row-wise in order of decreasing
  H$\alpha$ luminosity such that the top left thumbnail is the most
  luminous, and the bottom right is the least.
 There are two types of morphology: very compact or
 more extended (generally disturbed) systems. Many of the disturbed
 morphologies show evidence for two bright components within
 $\sim$ 1$''$ of each other -- indicative of recent or ongoing
 interaction which might be powering star-formation. The colour images
 appear to support this, with some of the more extended sources
 showing clear red and blue parts, reminiscent of merging star-forming
 galaxies in the local Universe.}
\end{figure*}

\subsubsection{Comparison to theoretical predictions of the luminosity
  distribution}

Recent recipes for galaxy formation using a semi-analytic
prescription provide us with the opportunity to directly compare observed
H$\alpha$ line luminosities with theoretical predictions (Bower et al\
2006). In Figure 5 we present the H$\alpha$ luminosity function at
$z=2.23$ from the Bower et al. (2006) model (hereafter `B06 model')
which populates dark matter halos in the $N$-body Milli-Millennium
simulation with model galaxies (1/16 the volume of the full Millennium simulation,
Springel et al.\ 2005). The B06 model is a variant of {\sc galform},
the Durham semi-analytic code (Cole et al.\ 2000; Benson et
al. 2003), which includes  a new prescription for AGN feedback.  Such
feedback appears to be very important for preventing the cooling of
baryons in massive galaxies, and thus steepening the bright end of the
luminosity function (see Bower et al. 2006). We take the {\it total}
H$\alpha$ luminosity from the catalogue, and enforce the same equivalent width
cut, $W_{\rm H\alpha} > 50$\AA as in our narrow-band selection. No
other constraint is placed on other observable parameters. 
The observed luminosity distribution generally appears to agree well
with the predictions, although there appears to be a slight excess
of bright H$\alpha$ emitters in the model. The origin of this
disparity could be caused by Cosmic variance effects
or over-production of bright star-forming galaxies in the
simulations, exacerbated by the low-number statistics in the
high-luminosity bins. Nevertheless, the overall rough agreement of the H$\alpha$
 luminosity density over the observed range is encouraging.

\subsection{The properties of H$\alpha$ emitters}

\subsubsection{Hubble Space Telescope ACS imaging -- morphologies}

In Figure~7 we present {\it Hubble Space Telescope
  (HST)} ACS F814W thumbnail images of the candidate
H$\alpha$ emitters. The exquisite imaging from  {\it HST} allows their
rest-frame UV morphologies to be examined, and we have
used this, along with {\it BRz} Subaru SuprimeCam optical imaging from the COSMOS
archive to preserve morphological information and reveal colour
gradients that vary on scales comparable to ground-based
imaging resolution. For example this allows us to distinguish red and
blue components in an interacting galaxy. We observe four 
morphological types: compact and isolated galaxies ($33\pm8$\%);
amorphous or disc
morphologies ($35\pm8$\%) and galaxies with obvious multiple components
($27\pm7$\%). Note that several of the galaxies
have too poor $S/N$ in the {\it HST} images to determine a
classification ($5\pm4$\%).  We note that the close-companion rate is only marginally
above that for a similar $K$-band  selected sample: 21\% of
$18<K<19.5$ galaxies selected from the full COSMOS survey have an
$I<25$\,mag neighbour within 2.5$''$.

The mean half-light radius ($r_{1/2}$, an effective measure of galaxies' size)
for the H$\alpha$ emitters is $\left<r_{1/2}\right> = 2.1\pm0.5$\,kpc.
How do these galaxies compare with other UV-selected
systems at the same epoch? Erb et al.\ (2004) report morphologies and sizes for
13 `BX' galaxies at $z\sim2$ in the GOODS-N field (although these
galaxies were selected partially for their known elongation, see also
F\"orster-Schreiber et al.\ 2006). The `BX'
galaxies appear to have qualitatively similar properties to many of
the H$\alpha$ selected galaxies in this work: clumpy, `tadpole'
features (see also Law et al. 2007). BX galaxies in the Erb
et al. (2004) sample have $r_{1/2} = 1.9\pm0.5$\,kpc, 
implying they have similar
sizes (and qualitatively similar morphologies) to H$\alpha$-selected star-forming
galaxies at the same redshift.

\subsubsection{{\it Spitzer} 24$\mu$m and VLA 1.4\,GHz observations --
  obscured star-formation}

The entire COSMOS field has been observed with the Multiband
Imaging Photometer (MIPS) on the {\it Spitzer Space Telescope}. The
24$\mu$m flux at $z\sim2.23$ probes polycyclic aromatic hydrocarbons
(PAH) emission lines. In dusty star-forming galaxies at this redshift,
24$\mu$m should be dominated by the prominent 7.7$\mu$m emission
feature, and so should
be a good indicator of obscured star-formation. To investigate
the possibility that some of the H$\alpha$ emitters are dusty luminous
starbursts, we cross-correlate the H$\alpha$ catalogue with the {\it
  Spitzer}-COSMOS (Sanders et al. 2006)
MIPS 24$\mu$m catalogues using a simple matching radius of 2$''$.
Seven H$\alpha$ emitters from our sample of 55 are
detected in both the wide and deep
24$\mu$m MIPS maps (flux limits of 0.3\,mJy and 0.06\,mJy respectively) with a
noise weighted average 24$\mu$m flux density of $\left<
  f_{\rm 24\mu m}\right> = 0.20\pm 0.03$\,mJy. 

In order to search for a statistical signature
of mid-infrared emission from the whole sample of H$\alpha$ emitters 
we stack the 24$\mu$m data in regions around the
source positions (the H$\alpha$
fluxes of 24$\mu$m detections are not significantly different from the
overall H$\alpha$ population, so we include these in the stack). The
resulting stacked image yields a significant flux excess,
$f_{\rm 24\mu m} = 0.11\pm 0.01$\,mJy. The uncertainty is derived by
stacking 50$\times$55 thumbnail images around random locations in the
24$\mu$m map, and measuring the scatter in the fluxes measured in
identical apertures in each stack. The significant 24$\mu$m flux suggest weak mid-infrared
emission in these star-forming galaxies. Since COSMOS was also
observed at 70$\mu$m and 160$\mu$m with MIPS, we
apply a similar stacking
analysis for these wavelengths in an attempt to place constraints on
the far-infrared properties of the H$\alpha$ emitters. We do not find
significant flux excess in either case, and so we define 3$\sigma$
upper limits of $f_{\rm 70\mu   m} < 2.1$\,mJy, and $f_{\rm 160\mu m}
<9.2$\,mJy.

How does the infrared
luminosity compare to the H$\alpha$ luminosity? To compare these two
values we first estimate the total-infrared luminosity by normalising an
M\,82-like spectral energy distribution to the observed 24$\mu$m flux
and integrating over 8--1000$\mu$m (the choice of SED is motivated by
the discussion in \S4.2.3). The SFR is estimated using the
calibration of Kennicutt  (1998), ${\rm SFR[IR]} =
4.5\times10^{-45}\,(L_{\rm 8-1000\mu m} / {\rm erg\,s^{-1}}
)\,M_\odot~{\rm yr}^{-1}$. We find ${\rm SFR[IR]} =
90\pm 40\,M_\odot$\,yr$^{-1}$, where the uncertainty is derived by
recalculating the total infrared luminosity by introducing a 50\%
variation in the normalisation of the SED, since at $z=2.23$ the
24$\mu$m flux probes the 7.7$\mu$m PAH emission line, and the fraction
that this band can contribute to the total infrared flux can vary.
How does this compare with the H$\alpha$ measurements? 
Without applying any correction for extinction at H$\alpha$, we
calculate the average
${\rm SFR[H\alpha]} = 40\pm20\,M_\odot$\,yr$^{-1}$ (the
uncertainty reflecting the 1$\sigma$ scatter in H$\alpha$
luminosities, and assuming ${\rm SFR[H\alpha]}
=7.9\times10^{-42}\,(L_{\rm H\alpha} / {\rm erg\,s^{-1}}
)\,M_\odot~{\rm yr}^{-1}$, Kennicutt et al. 1998). Including
 the canonical extinction correction of 1\,mag, this rises to  ${\rm SFR[H\alpha]} =
100\pm45\,M_\odot$\,yr$^{-1}$ in good agreement with the SFR
derived from the average infrared luminosity. As described above, the
true level of extinction could be a function of luminosity, however
this result suggests that our adoption of a constant reddening
correction does not result in a significant underestimate of the SFR.

Four H$\alpha$ emitters are formally detected in the 1.4\,GHz VLA map of
COSMOS (Schinnerer et al. 2004), with an
average flux of $\left< f_{\rm 1.4GHz}\right> = 160\pm
20$\,$\mu$Jy. Again, these objects' H$\alpha$ fluxes are not
significantly different from the general population, and so we employ
the same stacking technique to find the `average' 1.4\,GHz flux for
the sample. We find $f_{\rm 1.4GHz} =9.7\pm1.2$\,$\mu$Jy (the error is
calculated in the same manner as the 24$\mu$m stack -- i.e. repeating
the analysis for randomly co-added thumbnail images). Parameterising
the correlation between infrared and radio luminosity as $q_{24} =
\log(f_{\rm 24\mu m} / f_{\rm 1.4\,GHz})$  (see Ibar et al. 2008) we find
$q_{24} = 0.7\pm0.2$. This value is in good agreement with
 the expected $q_{24}$ for an M\,82 SED at $z=2.23$,
$q_{24}=0.6\pm0.1$  (Ibar et al. 2008) ,  and so converting the radio luminosity
to a far-infrared luminosity (and thus SFR) is consistent with our
derived SFRs from the 24$\mu$m luminosity and extinction corrected H$\alpha$
luminosity described above.

\subsubsection{Average spectral energy distribution of H$\alpha$
  emitters}

In Figure 8 we show the average spectral energy distribution of the
H$\alpha$ emitters. The optical/near-IR {\it uBVgrizK}+F814W+IRAC
3.6--8$\mu$m photometry is averaged over all the H$\alpha$ emitters
(weighted by the individual photometric errors) and then fit with a
choice of Sa-d and burst spectral templates (Bruzual \& Charlot 1993) redshifted to $z=2.23$
using the {\sc hyperz} photometric redshift code (Bolzonella et
al. 2000). Note that the IRAC measurements are the $3''$ aperture fluxes corrected to total flux
using the values suggested by the S-COSMOS team: 0.90, 0.90, 0.84 and
0.73 for the 3.6, 4.5, 5.8 and 8$\mu$m cameras respectively. Coverage of the rest-frame
near-infrared is useful for improving the photometric redshifts of
high redshift objects, since it allows the location of the 1.6$\mu$m
stellar bump to be tracked (e.g. Sawicki 2002). 
The wavelength of this feature corresponds to the minimum
opacity of H$^-$ ions in the atmospheres of cool stars (John 1988), and thus is an
indicator of the evolved ($\gtrsim20$\,Myr old) stellar population.    
The average 3.6--8$\mu$m SED agrees with
the 1.6$\mu$m feature redshifted to $z=2.23$.  
The data are well fit by an Sc spectral type template with a
foreground screen 
reddening of $A_V = 1.0$\,mag (this parameter was free to vary in the
range $0 < A_V < 2$). This suggests the UV continuum in these sources
may be slightly more obscured than we are assuming for the H$\alpha$ line
emission. 

The good fit to the spectral template is compelling evidence 
that our narrow-band plus photometric selection is indeed identifying
a population of $z=2.23$ star-forming galaxies, and shows that improved photometric
redshifts that make use of bands redward of $K$ will help in improving
the narrow-band selection (Mobasher et al. 2007). 
As IRAC is probing rest-frame near-infrared emission, 
we can estimate the stellar mass of these galaxies from the absolute
rest-frame $K$-band luminosity. We estimate the rest-frame absolute $K$-band
magnitude from the 5.8$\mu$m flux, $M_K = -25.4$\,mag (AB
magnitude).  The stellar mass, $M_\star$, is estimated using a similar
method to Borys et al. (2005), assuming $M_\star$ is equal to the
integrated stellar mass at the age of the galaxy. Note that our
template assumes a continuous star-formation history, with a Miller \&
Scalo (1979) initial mass function (a burst template does not provide
a satisfactory fit to the observed photometry). If the $K$-band luminosity traces
stellar mass, then $M_\star = 10^{-0.4(M_K-3.3)} /(L_K/M)\,M_\odot$
where $L_K/M$ is the mass-to-light ratio, and the factor 3.3
corresponds to the absolute solar $K$-band luminosity (Cox\ 2000). For
continuous star-formation we estimate $L_K/M$ from the description of
Borys et al.\ (2005) who use the {\sc Starburst99} stellar synthesis
code (Leitherer et al.\ 1999) to find $L_K/M = 103(\tau/{\rm
  Myr})^{-0.48}$ for ages $\tau >100\,{\rm Myr}$. The age returned
from {\sc hyperz} is 1\,Gyr, similar to the mean age of $z\sim2$
UV-selected star-forming galaxies in the survey of Erb et
al. (2006). Adopting a realistic range in ages, 0.5--2\,Gyr we find 
$M_\star\sim6.1$--$11.5\times10^{10}M_\odot$.  This is larger than
the mean stellar mass of the sample in Erb et al. (2006), who find $M_\star =
(3.6\pm0.4)\times10^{10}\,M_\odot$, but the disparity between the two
mass estimates may be simply due to the different derivation methods used:
comparison of the spectral energy distributions to stellar synthesis
models in Erb et al. (2006) versus simple age-dependent mass-to-light ratios here.
     
In addition to the UV--near-IR photometry, we also show on Figure 8 the
longer wavelength data described above: 24$\mu$m, 70$\mu$m, 160$\mu$m and 1.4,GHz\,cm
statistical detections and limits. 
For comparison we show the M\,82 galaxy SED (a local starburst)
redshifted to $z=2.23$ and scaled in flux to match the 24$\mu$m flux
density accordingly. 
The average 24$\mu$m/1.4\,GHz flux is consistent with what would be
expected for a typical star-forming galaxy at $z=2.23$ (Boyle et
al. 2007; Ibar et al. 2008).
This suggests that on average these galaxies are unremarkable
starbursts, with SFRs $\lesssim100$\,$M_\odot$\,yr$^{-1}$, with
properties broadly similar to the most actively star-forming galaxies
found in the local Universe. We note however, that there is a departure
from the similarity to M\,82 in the rest-frame UV, with our composite
SED showing quite significant UV-excess over what would be expected
from M\,82 at this redshift (Ibar et al. 2008). We interpret this as indicating that
perhaps on average, the observed H$\alpha$ emitters are not as dusty as
the local LIRG, or they simply have higher escape fractions of UV photons.

\begin{figure}
\includegraphics[width=3.4in]{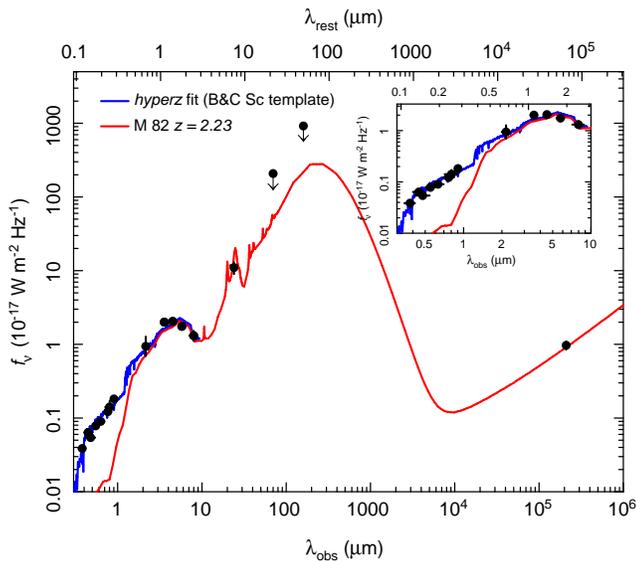}
\caption[Average spectral energy distribution of H$\alpha$
emitters]{The average SED for H$\alpha$ emitters based on
  UV--radio photometry from the COSMOS archive averaged over
  all  55 candidates. In the inset we fit these with a Sc galaxy
  template at $z=2.23$ reddened
  by $A_V = 1$\,mag.  The fit agrees well with the average SED,
  suggesting that our narrow-band selection is indeed identifying star
  forming galaxies at $z=2.23$.
 The main panel extends the SED to
  longer wavelengths by stacking the 24$\mu$m, 70$\mu$m, 160$\mu$m  and 1.4\,GHz
  data (the 70$\mu$m and 160$\mu$m points are 3$\sigma$ upper-limits), and we
  overlay the M\,82 template, redshifted to $z=2.23$ and
  scaled in flux accordingly. Again, this seems to verify the redshift
selection, and suggests that the galaxies we select are similar to
local starburst galaxies in their broad SEDs.}
\end{figure}

\begin{figure}
\centerline{\includegraphics[width=3.25in]{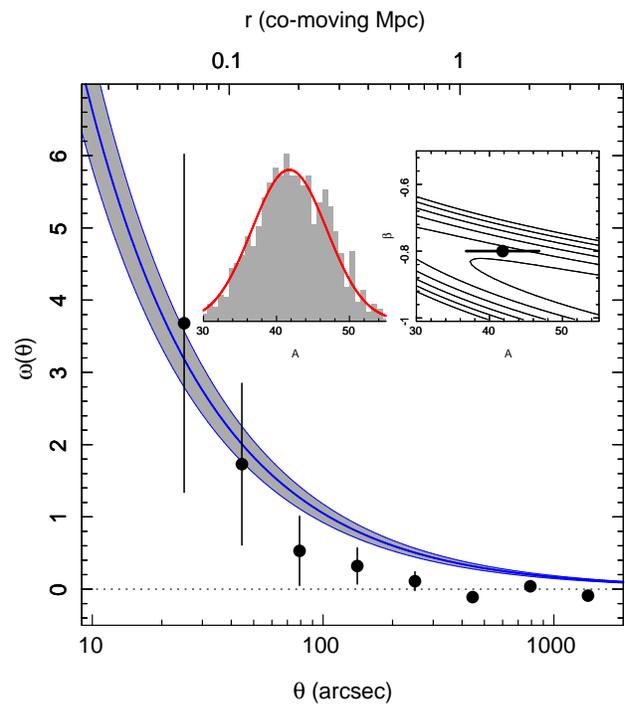}}
\caption{The two-point angular correlation function of H$\alpha$
  emitters, $\omega(\theta)$.  
The data are fit with a power law $A\theta^{-0.8}$, and
  we find $A = 41.8\pm 5.0$ (for $\theta$ in arcseconds), corresponding to a
correlation length $r_0 = 4.2^{+0.4}_{-0.2}\,h^{-1}$\,Mpc. The insets
  show the likely uncertainty on the fit. The histogram shows the
  distribution of amplitudes when $\omega(\theta)$ is recalculated for
  a different realisation of the random catalogue (this was repeated 1000
  times). The contour plot represents $\chi^2$ when the fit is allowed
  to vary both the amplitude and slope $\beta$ of the power-law fit,
  thus demonstrating the covariance dependence on the uncertainty in
  $A$. We find an adequate fit is obtained without varying $\beta$,
  and therefore in our estimates we fix it at the fiducial value of
  $\beta = -0.8$, but it appears a better fit would be found for
  $\beta\sim-1$, and a larger amplitude. The effect of this increase
  in amplitude and $\beta$ would correspond to a slight decrease in $r_0$:
  $A=95$, $\beta=-1$ corresponds to $r_0\sim3.6\,h^{-1}$\,Mpc; in
  better agreement with the measurement from the simulation (\S4.3.2).
  Thus our current estimate of the clustering strength might be a
  slight overestimate. These are preliminary findings -- a better estimate
  requires a larger sample. However, these initial results suggest
  moderate clustering on scales of 10--100$''$ (i.e. sub-Mpc or sub-halo).}
\end{figure}

\subsection{Clustering of H$\alpha$ emitters}

\subsubsection{Evaluating the clustering strength}

Estimating the clustering properties of star-forming galaxies can be a
powerful tool in understanding their formation and evolution, because
the clustering strength provides information about the dark matter
halos these galaxies reside in. The descendants (and progenitors) of
these galaxies can then be traced by analysing the evolution of the
halos themselves.  
Despite our relatively small sample size, we note that it is selected
in a relatively narrow redshift range with little dilution of the
projected clustering, so we attempt to estimate the clustering
strength of H$\alpha$ emitters using the
angular correlation function, $\omega(\theta)$, from which we can
derive the physical correlation length, $r_0$. We use the estimator
proposed by Landy \& Szalay (1993):  
\begin{equation}\omega(\theta)  =
  1+\left(\frac{N_R}{N_D}\right)^2\frac{DD(\theta)}{RR(\theta)} -
  2\frac{N_R}{N_D}\frac{DR(\theta)}{RR(\theta)}.\end{equation} Here
$DD(\theta)$ is the number of pairs of real `data'
galaxies within $(\theta,\theta+\delta\theta)$, $DR(\theta)$ is the
number of data-random pairs and $RR(\theta)$ is the number of
random-random pairs. $N_R$ and $N_D$ are the number of random and data
galaxies in the survey. The random catalogue is generated by uniformly
distributing $20N_D$ false galaxies over a geometry corresponding to the
survey field-of-view, which is appropriate for the small sample. 
The errors for each angular interval (spaced at 0.25 dex) are
estimated with $\delta\omega(\theta) = (1+\omega(\theta)) /
\sqrt{DD(\theta)}$, which estimates the Poisson noise in each bin
(Landy \& Szalay 1993).  
We present our results in Figure~9, and fit
the data with a power-law of the form $A\theta^{-0.8}$, where $A$ is
the amplitude of the angular correlation function, and the power-law
index $-0.8$ is the fiducial value. For $\theta$ in arcseconds, we find
$A=41.8\pm 5.0$. The uncertainty is the 1$\sigma$ range of $A$ when the fit is
repeated for 1000 realisations of $\omega(\theta)$, with different
random catalogues. The integral constraint, $C$, which accounts for
bias introduced by  only observing a small
region of the sky, will result in a reduction of the amplitude of
the two point correlation function: $A(\theta^{-0.8} - C)$. 
We estimate the effect following
Roche et al.\ (2002), and find it to be small for our survey area, with
$C = 0.0055$ for $\theta$ in arcseconds -- we include it when fitting
the amplitude $A$.

A more useful description of the physical clustering of galaxies is the
real-space correlation length, $r_0$, and this can be calculated directly from
the angular correlation function using the inverse Limber
transformation (Peebles 1980; Efstathiou et al.\ 1991) which relates the
spatial correlation function to the angular correlation function
(provided the redshift distribution is known). 

 Using clustering simulations described below, we have 
modeled the effect of the Gaussian
filter selection function on our clustering measurements.
We compare the projected clustering amplitude, $A$,
derived for the population in  a uniformly selected volume (i.e.\
assuming
a top-hat transmission function) 
to that derived for samples selected
with a luminosity-dependent selection function 
appropriate for our filter transmission curve.
We find no statistical difference when
including a luminosity-dependent volume correction, with $\left<A_{\rm
    Gaussian}\right> \simeq \left<A_{\rm
    uniform}\right> \sim 45$, using 100 simulations for each
method. Therefore, for simplicity 
in the following  deprojection analysis 
we treat the projected correlation amplitude
as if it came from a uniformly selected sample over the narrow redshift slice at
$z=2.22$--2.25.
We also assume that the real space correlation
function is independent of redshift over this narrow range (this is
the same approach used in other high redshift narrow-band surveys, e.g.\
Kova\v c et al.\ 2007). 

We find $r_0 = 4.2^{+0.4}_{-0.2}\,h^{-1}$\,Mpc, where the
uncertainty is derived by reevaluating $r_0$ after applying the
1-$\sigma$ uncertainty on the fit to $\omega(\theta)$ described
above. This likely underestimates the true error due to added effects
such as cosmic variance, which the reader should note could be an
important effect in our $220\times10^3$\,Mpc$^3$ survey
and to assess this we now turn to our simulations.

\subsubsection{Comparison to theoretical predictions of the clustering
  properties of star-forming galaxies}

In addition to the luminosity distribution predicted by semi-analytic
recipes (\S 4.2.2), we can also compare observed clustering properties of
H$\alpha$ emitters to theoretical predictions from these same models. 
The B06 model populates dark matter
halos within a $\Lambda$CDM framework cosmology, where the
`observable' properties such as optical luminosity are dictated by a
physically motivated prescription for the behaviour of baryons cooling
within the halos. Thus, we can directly evaluate $r_0$ for H$\alpha$
emitters within the simulation volume.

Taking the B06 model,  we limit simulated luminosities to the range
$42 \leq \log L_{\rm H\alpha}~{\rm
  (erg~s^{-1})}~\leq44$ (approximately the range of observed
luminosities) and $W_{\rm H\alpha} > 50$\AA, and then project the
positions of the galaxies onto a 2D `sky'. We then measure the
amplitude of the two-point angular correlation function and convert to
$r_0$ in exactly the same way as before (\S4.3.1). We find $r_0 =
5.7^{+0.6}_{-0.7}$\,$h^{-1}$\,Mpc, slightly larger than our observed value
(but within $\sim$2$\sigma$). Part of this disparity could be due to the
slight excess of bright H$\alpha$ emitters in the simulations: the
clustering of H$\alpha$ emitters in the $z=0.24$ study of Shioya et
al. (2008) was seen to increase with $L_{\rm H\alpha}$.  The reader
should also take into account the relatively small size of our survey (despite it being the
largest to date) compared to the effective area of the
Milli-Millennium. Thus, there are additional uncertainties on our estimate of
$r_0$ due to the likely  field-to-field variations which are still
significant on this scale.

One way of illustrating this is to randomly select sub-volumes in the
Milli-Millennium box corresponding to our survey volume. Projecting
galaxies selected in these sub-volumes allows us to evaluate the
amplitude of their angular correlation function in an identical way to that
performed in \S4.3.1. The scatter in amplitude (fixing the power-law slope to the
fiducial value, $-0.8$, as before) for a set of random positions is a good
indicator of the expected Cosmic variance on these
scales. Statistically we find a similar amplitude to the real data,
$\left< A \right> \sim 45\pm31$ (for $\theta$ in arcseconds), but with large
scatter from the field-to-field variations.  
Thus, the dominant source of uncertainty in our measurement
of the clustering strength of these objects is Cosmic variance,
exacerbated by the relatively small numbers of objects in our sample.

  \begin{figure}
\includegraphics[width=3.3in]{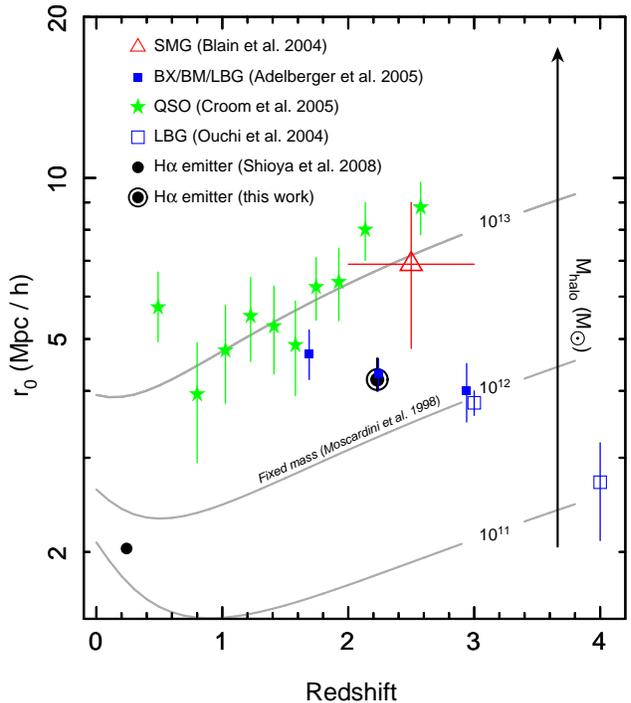}
\caption{Evolution of the correlation length $r_0$ in co-moving Mpc
  for a variety of galaxy populations over $z=0$--4. Literature points
  include submm galaxies (Blain et al. 2004); QSOs (Croom et al. 2005);
  LBGs (Ouich et al. 2004; 2005); BM/BX
  galaxies (Adelberger et al. 2005), and local H$\alpha$ selected
  galaxies (Shioya et al. 2008).
We relate the clustering properties to the expected clustering from
  models of the clustering of dark matter halos, assuming a model for
  the evolution of bias (Moscardini et al. 1998). Our results suggest
  that H$\alpha$ emitters at $z=2.23$ are hosted by dark matter halos
  of mass $\sim$$10^{12}\,M_\odot$ -- in good agreement with the masses of
  halos expected to host similarly selected galaxies at the same
  redshift (i.e. BM/BX galaxies). 
Thus, these H$\alpha$ emitters appear to be the progenitors
  of Milky Way-like galaxies at $z=0$.}
\end{figure}

\subsubsection{Halo mass}

The bias parameter describes how the observed galaxy distribution
traces the underlying matter distribution. Assuming a model for the
evolution of bias over cosmic time, it is possible to estimate the
host halo mass of different galaxy populations by comparing their
correlation length, $r_0$ to that of galaxies within halos of various
mass. More importantly, this analysis can provide clues to determine
the progenitor populations of galaxy populations in the local
Universe. In Figure 10, we compare our results to predictions of $r_0(z)$ for
dark matter halos with a fixed {\it minimum} mass dark matter halos
of mass $M_{\rm min} > 10^{11-13}\,h^{-1}\,M_\odot$ (Matarrese et
al. 1997; Moscardini et al. 1998). This has become a popular way to
present the results of clustering analyses in a cosmological context 
(e.g. Overzier et al. 2003; Blain et al. 2004; Farrah et al. 2006). 
We assume a $\Lambda$CDM cosmology, and an evolving bias model $b(z)$
(Moscardini et al.\ 1998). We use the values
tabulated by Moscardini et al. (1998) for various fixed minimum mass
halos. We evaluate the predicted $r_0$ of galaxies by solving 
$\xi_{\rm gal}(z) = D(z)^2b(z)^2\xi_{\rm
    DM}(0)$, where $D(z)$ is the growth factor and $\xi$ is the real
  space correlation function. 

Our results suggest that H$\alpha$ emitters at $z=2.23$ reside in
moderate mass halos, with $M_{\rm min}$ of order
$10^{12}\,M_\odot$, similar to that of a Milky Way mass halo at
$z=0$, suggesting that these high redshift star-forming galaxies are
the progenitors of $M_\star$ galaxies seen today.
Given that the semi-analytic model is tied directly to the underlying
dark matter halos it is
possible to extract the halo masses that the B06 H$\alpha$ emitters
actually reside in. Limiting the model luminosities to the observed range,
we find halo masses spanning $\log M_{\rm halo}\,(M_\odot)  =
11.2-12.9$. with a median of $\log M_{\rm halo}\,(M_\odot)  = 11.7$. This
is in reasonable agreement with the halo mass estimate from the
observed deprojected angular correlation function.

In comparison to other active populations at the same epoch (QSOs,
SMGs and UV-selected BX/BM galaxies), we see that H$\alpha$ emitters
typically reside in lower-mass halos than SMGs and QSOs and are not
directly related to them, but the H$\alpha$ emitters have a
similar clustering strength to BX/BMs at the same redshift
(Fig~10). This is perhaps not surprising, given the overlap in
selection between the two populations (Fig.~3), and suggests moderate
clustering of star-forming galaxies at this redshift.

\section{Summary}

We have presented results from the largest near-infrared narrow-band
survey for H$\alpha$ emission at $z=2.23$ yet undertaken, probing
$220\times10^3$\,co-moving Mpc$^3$ to a H$\alpha$ line flux of
$1\times 10^{-16}$\,erg\,s$^{-1}$\,cm$^{-2}$. Selecting potential
$z=2.23$ H$\alpha$ emitters on the basis of the significance of their
narrow-band excess at 2.121$\mu$m and broad-band colours and magnitudes,
we detect 55 galaxies over 0.603\,square degrees in the COSMOS field,
a volume density of $(2.5\pm0.3)\times10^{-4}$\,Mpc$^{-3}$. Our
findings can be summarised as follows:

\begin{itemize}

\item{The luminosity function is well fit by a Schechter function with
     $\phi^\star=(1.45\pm0.47)\times10^{-3}$\,Mpc$^{-3}$, $\log
     L^\star = {42.83\pm0.13}$\,erg\,s$^{-1}$ assuming a faint end
    slope of $\alpha=-1.35$, as found by Gallego et al. (1995) for the
    H$\alpha$ LF at $z=0$. We find strong evolution in the H$\alpha$
    luminosity function from $z=0$--2.23 equivalent to a 10$\times$
    increase in characteristic luminosity. The $z=2.23$ LF is very similar to the $z=1.3$
    distribution of Yan et al. (1999), implying little evolution in
    the population between these epochs. We note however that deeper
     observations will be required to probe the faint end of the LF,
     which could impact on our estimate of the integrated $\rho_{\rm SFR}$.}

\item{The extinction corrected SFRD determined from the integrated LF is
    $0.17^{+0.16}_{-0.09}$\,$M_\odot\,{\rm yr^{-1}}\,{\rm Mpc^{-3}}$
    for $L_{\rm H\alpha} > 10^{42}$\,erg\,s$^{-1}$ $z=2.23$ (including
    a 15\% correction for AGN contamination to the luminosity
    density). This
    result for the first time directly traces the evolution of $\rho_{\rm SFR}$ out to
    $z=2.23$ using H$\alpha$ alone, and supports the view that the SFRD
    undergoes a plateau at $z\sim$1--2 prior to sharp decline to the
    present day.}

\item{The H$\alpha$ emitters have infrared and radio properties
    similar to that expected for local star-forming galaxies: our
    statistical measurement as of the 24$\mu$m and 21\,cm flux
    densities are consistent with a LIRG SED at $z=2.23$, with the infrared
  derived SFR agreeing with the H$\alpha$ derived rate 
  if we assume the canonical H$\alpha$ extinction, $A_{\rm
    H\alpha}=1$\,mag. However, in comparison to M\,82 (an archetypal
    LIRG), we observe a UV light excess over the
    template, which suggests that these galaxies are not as obscured as
    M\,82, or at least the escape efficiency of UV photons is higher
    in these galaxies. {\it HST}-{ACS} imaging reveals that the
H$\alpha$ emitters have a range of morphologies including compact and
disturbed systems: several comprise of two components separated on
scales of $\lesssim$10$''$\,kpc, indicating recent or ongoing merger
activity, as might be expected for starburst galaxies. Quantitatively
the rest-frame UV-morphologies of the H$\alpha$ emitters are similar
to UV-selected star-forming (`BX') galaxies at the same redshift, with
mean half-light radii of $r_{1/2} = 2.1\pm0.5$\,kpc. 
}

\item{We have performed the first clustering analysis of
H$\alpha$ emitters at $z=2.23$. First, we measure the amplitude of the
two-point angular correlation function $\omega(\theta)$, and find $A =
41.8\pm5.0$ for $\theta$ in arcseconds. The inverse Limber transform
provides us with the real-space correlation length $r_0$, which we find
is $r_0= 4.2^{+0.4}_{-0.2}$\,$h^{-1}$\,Mpc. Comparing this result to
other galaxy populations over cosmic time, these high redshift star
forming galaxies have similar clustering properties to similarly
selected galaxies at high and low redshift, and reside in dark matter
halos of masses of $\sim$$10^{12}\,M_\odot$. This is agrees with 
the range of masses of host halos of H$\alpha$ emitters selected from
the Millennium simulation. Our H$\alpha$ emitters appear to be slightly less
strongly clustered than QSOs and SMGs, but similar to BX/BM galaxies at this
redshift. Comparisons to models of
clustering evolution taking into account bias evolution, our results
suggest that these galaxies will likely evolve into
$L^\star$ galaxies, similar to the Milky Way at $z=0$. }
\end{itemize}

These observations represent the first results of 
an innovative survey for
emission line objects over several square degrees at $z=0.84$, 1.47
and 2.23 corresponding to narrow-bands in $J$, $H$ and $K$: the Hi-Z Emission
Line Survey (HiZELS) on UKIRT.
The outcome of this survey will be large samples ($\sim$1000 at each redshift) of identically
selected `typical' star-forming galaxies within well defined volumes. 
The large area will enable us to place even stronger constraints on the star
formation history over this crucial epoch from
identically selected star-forming galaxies in each redshift slice. Our
custom-made $J$ and $H$ narrow-band filters at 1.204$\mu$m and 1.618$\mu$m
respectively will not only detect H$\alpha$ emission at $z=0.84$ and
$z=1.47$, but also [O~{\sc ii}]$\lambda3727$  and  [O~{\sc
  iii}]$\lambda5007$ emission lines from sources at $z=2.23$, exactly
the same redshift as the current 2.121$\mu$m survey. This will be a
useful test for confirming
the redshifts of a subset $z=2.23$ H$\alpha$ emitters by matching to
detections in these other narrow-band filters. It will also provide a
more complete survey of the emission line population of star-forming
galaxies and AGN at this epoch.

\section*{Acknowledgements}

We thank the referee for a careful reading of the manuscript, and
appreciate several helpful suggestions that have improved the quality
of this work. JEG \& KC thank the U.K. Science and Technology
Research Facility (STFC, formerly PPARC) for financial support, IS \&
PNB acknowledge the Royal Society. 
JK thanks the German Science Foundation for
financial support by way of SFB439. The authors wish to thank Calton
Baugh, Richard
Bower, Gavin Dalton, Helmut Dannerbauer, Alastair Edge, Vince Eke, Duncan Farrah,
John Helly, Cedric Lacey,
Simon Lilly, Bahram Mobasher, Marco Riello, Tom Shanks and David Wake
for helpful discussions. Finally, it is a pleasure to thank the
support astronomers and telescope support staff at JAC/UKIRT who 
provided an excellent service throughout the observations required to
complete this work: Andy Adamson, Luca Rizzi, Tom Kerr, Thor Wold, Tim
Carroll and Jack Ehle.  UKIRT is funded by the STFC.

\setlength{\bibhang}{0.25in}

\end{document}